\documentclass[%
 aip,
 sd,%
 amsmath,amssymb,
 reprint,%
]{revtex4-1}

\usepackage{graphicx}
\usepackage{dcolumn}
\usepackage{bm}

\begin{document}

\preprint{AIP/123-QED}

\title{Energy evolution in  many-particle quantum hydrodynamics of spinning plasmas}
\author{Mariya Iv. Trukhanova}
\email{mar-tiv@yandex.ru}

\affiliation{Faculty of physics, Lomonosov Moscow State University, Moscow, Russian Federation.}

\date{\today}%

\begin{abstract}
In this paper, we develop a quantum hydrodynamics (QHD) method for the research of
the quantum evolution of a system of spinning particles.
 We derived the fundamental equation for charged and neutral spinning particles  -  the energy evolution equation  from the many-particle microscopic
Schredinger equation with a \emph{Spin-spin} and \emph{Coulomb} modified Hamiltonian. We derive the spin contributions to the energy evolution equation, thermal energy and thermal energy current.
\end{abstract}

  \maketitle

\section{Introduction}
The properties of  quantum plasmas had
been studied by several authors \cite{1} - \cite{7}.          The quantum and magnetization effects  can be produced by two main
 terms. The first is a quantum force, the  quantum \emph{Bohm} potential,  is proportional to $\hbar$ and has a purely quantum origin \cite{111}, \cite{112}.  The second force  given by action of
magnetic field on magnetic moments and the magnetic moment evolution equation. \cite{1}. The complete theory of spin-1/2 electron-positron quantum plasmas had been discussed in Ref. \cite{61}, where  the Coulomb, spin-spin, Darwin and annihilation interactions had been taken into account.   The set of   quantum   hydrodynamics equations for charged and neutral  spinning particle using the Pauli equation  had been performed in  Refs. of T. Takabayashi and P. Holland   \cite{A8} - \cite{A110}.  In Ref. \cite{A8} the author derived a set of hydrodynamic equations for a single spinning particle using the vector representation of a spinning particle and found the non-linear dynamical terms in the equations of motion. These terms have been interpreted as the \emph{"internal potential"} and \emph{"internal magnetic field"}.

Processes involving $N$
interacting particles are revealed in three dimensional physical
space, and are not very pronounce in the 3N dimensional configuration space. As wave
processes, processes of information transfer, diffusion and
spin transport processes occur in the three-dimensional
physical space. There are mathematical methods describing systems in three dimensional
space, such as hydrodynamics  \cite{1}, \cite{2}. The method of the many-particles quantum hydrodynamics  presents a mechanism to
project coordinate all particles from 3N dimension configuration space to 3D physical space
during derivation of equations of collective motion for application to coherent many-particle
structures.

Two kinds of the quantum electrodynamic radiative corrections to electromagnetic interaction and their influence on properties of highly dense quantum plasmas had been  considered in Ref. \cite{1112}.
Kinetic equations for the systems of spinning neutral particles with long-range anisotropic interparticle interaction were derived by a new method, which is the generalization of the many-particle quantum hydrodynamics. Spin waves in three- and two dimensional systems of neutral spinning particles had been investigated in Ref. \cite{A1}.
Many particle quantum hydrodynamics based on the Darwin Hamiltonian (the Hamiltonian corresponding to the Darwin Lagrangian) had been considered in Ref. \cite{A2}. A force field appearing in corresponding Euler equation had been considered in details and the semi-relativistic generalization of the quantum Bohm potential was obtained. The definition of the spin-current and the equation of the spin-current evolution had been derived earlier  \cite{A5}. The spin current equation derived from the many-particle microscopic Schrodinger equation was used to study the dispersion of collective excitations in three dimensional samples of magnetized dielectrics.

The QHD model, which separately describes spin-up electrons and spin-down electrons  had been derived in Refs. \cite{25}, \cite{26}. Applying the separated spin evolution quantum hydrodynamics to two-dimensional electron gas in plane samples and nanotubes located in external magnetic fields the authors find new kind of wave in electron gas, which is called the spin-electron acoustic wave \cite{26}.

The quantum hydrodynamic  with the Coulomb exchange interaction for three and two dimensional plasmas had been developed in \cite{27} and microscopic derivation of the Coulomb exchange interaction for electrons located on the nanotubes was  presented in \cite{A3}. The non-linear Schrodinger equations (NLSEs) for the Coulomb quantum plasmas with the exchange interaction had been derived and contribution of the exchange interaction in the dispersion of the Langmuir, and ion-acoustic waves was founded. The numerical simulation of the dispersion properties of different types of waves in the magnetized quantum plasmas with the Coulomb exchange interaction had been obtained in Ref. \cite{28}. The authors focused the attention to the Langmuir waves, Trivelpiece-Gould waves, ion-acoustic waves in non-isothermal magnetized plasmas, the dispersion of the longitudinal low-frequency ion-acoustic waves and low-frequencies electromagnetic waves.

Energy evolution equation for spinless  particles with the Coulomb interaction had been derived in Ref. \cite{A4}. The  dispersion relation
for Langmuir waves in quasi-neutral system of electrons and immobile ions using the energy evolution equation  was obtained.  In this article for studying of energy evolution and spin vortex effects we  generalize and use the method of the
many-particle quantum hydrodynamics  (MQHD) approach.  We propose a method
of quantum hydrodynamics that allows one to obtain a description of the collective effects in magnetized
quantum plasmas in terms of functions in physical space.
 We derive the fundamental balance equation, the fundamental energy evolution equation,  the magnetic moment evolution equation.    The fact that the thermal spin fluctuations were taken into account.


\section{\label{SS1}General Theory}

  The non-relativistic evolution of spinning particles represented by its spinor $\psi_s$  and given by     the  many-particle Schrodinger equation

  \begin{equation}
            i\hbar\partial_t\psi_s(R,t)=(\hat{H}\psi)_s(R,t),
  \end{equation}
where $R=(\textbf{r}_{1},...,\textbf{r}_{N})$.  A state of the system of $N$ particles is determined by the wave function in the 3N-dimensional configuration space, which is a $rank-N spinor$
            \begin{equation}
            \psi_s(R,t)=\psi_{s_1,s_2,....s_N}(\textbf{r}_{1},...,\textbf{r}_{N},t).
  \end{equation}
 Microscopic Hamiltonian for spinning particles has a form of

              \begin{equation} \label{H}    \hat{H}=\sum^{N}_{j=1}({\frac{\hat{D}^{2}_{j}}{2m_j}+q_j\varphi_{j,ext}-\mu_j\hat{\sigma}^{\alpha}_{j}B^{\alpha}_{j,ext}})
\end{equation}
\[
\qquad\qquad+\frac{1}{2}\sum^{N}_{j\neq k} q_jq_kG_{jk}-\frac{1}{2}\sum^{N}_{j\neq k,k}\mu^{2}_{j}F^{\alpha\beta}_{jk}\hat{\sigma}^{\alpha}_{j}\hat{\sigma}^{\beta}_{k},
\]
\[
\]
where the following designations are used in the (\ref{H}): $m_j$ - is the mass, $q_j$ stands for the charge of electrons $q_e=-e$ or for the charge of positrons $q_p=e$, $\mu_j=g\mu_B$, $\mu_{jB}$ - is the  magnetic moment  and $\mu_{jB}=q_j\hbar/2m_jc$ - is the gyromagnetic ratio
for electrons  and $\hbar$ - is
the Planck constant,  $g=1+\alpha/2\pi$ and $\alpha=1/137$ - is the fine structure constant, gets into account
the anomalous magnetic moment of electron. We introduce that $\hat{D}^{\alpha}_j=-i\hbar\hat{\nabla}^{\alpha}_j-q_jA^{\alpha}_j/c$ and $\textbf{A}_{ext}, \varphi_{j,ext} $  -  are   the vector and scalar potentials of
external electromagnetic field. the Green's functions  We also use of  the $Coulomb$ and $Spin - Spin$ interaction as $G_{jk}=1/r_{jk},$   $F^{\alpha\beta}_{jk}=4\pi\delta_{\alpha\beta}\delta(\textbf{r}_{jk})+\partial^{\alpha}_{j}\partial^{\beta}_{k}(1/r_{jk}).$

The first step in the        construction of MQHD apparatus is
to determine the concentration of particles $n(\textbf{r},t)$ in the neighborhood
of $\textbf{r}$ in a physical space

                               \begin{equation}
n(\textbf{r},t)=\sum_{s}\int dR\sum^{N}_{j}\delta(\textbf{r}-\textbf{r}_{j})\psi^{+}_s(R,t)\psi_s(R,t),
\end{equation}

\subsection{\label{sec:level2}Velocity field}

The velocity of $j-th$ particle $\textbf{v}_j$ is determined by
equation

\begin{equation} \label{z4}
         \textbf{v}_j=\frac{1}{m_j}(\nabla_jS-i\hbar\varphi^+\nabla_j\varphi)-\frac{q_j}{m_j c}\textbf{A}_j,
           \end{equation} The quantity $\textbf{v}_j(R,t)$ describes the current of probability
connected with the motion of $j-th$ particle, in general case
$\textbf{v}_j(R,t)$ depend on the coordinates of all particles of the system
$R$, where $R$ is the totality of $3N$ coordinate of $N$ particles
of the system $R=(\textbf{r}_{1},...,\textbf{r}_{N})$.

  The $S(R,t)$ value in the formula (\ref{z4}) represents the phase
of the wave function and as
the electron has spin, the wave function is now be expressed in the form o
\begin{equation}\label{psi}
\psi_s(R,t)=a(R,t)e^{\frac{iS}{\hbar}}\varphi_s(R,t),
\end{equation}
   where $\varphi_s$, normalized such that $\varphi^+_s\varphi_s=1$, is the new spinor, defined in the local frame of reference
with the origin at the point $\textbf{r}$. The spinor $\varphi_s$ gives the spin part of the wave function \cite{12}.

   As the particles of the system under consideration interact
via long-range forces the approximation of a selfconsistent
field is sufficient to analyze collective processes.
With the use of this approximation two-particle functions
in the momentum balance equation can be split into a product
of single-particle functions.  Taken in the approximation
of self-consistent field and the definition (\ref{z4}), the set of QHD equation has a form    \cite{12}

                         \emph{continuity
equation}
                               \begin{equation} \label{nnn}
\partial_{t}n+\nabla(n\textbf{v})=0,
\end{equation}

       \emph{momentum balance equation}
             \begin{equation} \label{jjjjj}
mn(\partial_{t}+\upsilon^{\beta}\partial_{\beta})\textbf{v}=qn\textbf{E}_{ext}+\frac{q}{c}n\textbf{v}_{e}\times\textbf{B}_{ext}-\nabla\wp\end{equation}
\[+\frac{\hbar^2}{2m}n\nabla(\frac{\triangle\sqrt{n}}{\sqrt{n}}) +M_{\beta}\nabla B^{\beta}_{ext}+\frac{\hbar^2}{4m\mu^2}\partial_{\beta}\Biggl( M^{\gamma}\nabla\partial^{\beta}(\frac{M^{\gamma}}{n})\Biggr)        \]

\[-n\nabla\int d\textbf{r}^{'}q^2G(\textbf{r},\textbf{r}^{'})n(\textbf{r}^{'},t)+M_{\gamma}\nabla\int d\textbf{r}^{'}F^{\gamma\delta}(\textbf{r},\textbf{r}^{'})M^{\delta}(\textbf{r}^{'},t),
  \]

            \emph{magnetic moment density equation}

   $$
(\partial_{t}+\upsilon^{\beta}\partial_{\beta})\textbf{M}=\frac{2\mu}{\hbar}\textbf{M}\times\textbf{B}_{ext} +\frac{\hbar}{2m\mu}\partial_{k}\Biggl(\textbf{M}\times\partial^{k}(\frac{\textbf{M}}{n})\Biggr)
 $$

 \begin{equation} \label{Mm}\qquad\qquad+\frac{2\mu}{\hbar}\epsilon^{\alpha\beta\gamma}M^{\beta}\int d\textbf{r}^{'}F^{\gamma\delta}(\textbf{r},\textbf{r}^{'})M^{\delta}(\textbf{r}^{'},t),
\end{equation} Let痴 discuss the physical significance of terms on the
right side of (\ref{nnn}) - (\ref{Mm}).    The first and second terms on the  right-hand side of Eq. (\ref{jjjjj}) characterize  the influence of the external electromagnetic field. The first term  on the
right side of Eq. (\ref{jjjjj}) describes  the action
of the external electric field $\textbf{E}_{ext}$ on the charge density $n$ and the second term on the
right side of Eq. (\ref{jjjjj})  represents  the Lorentz force field.  The third term on the  right-hand side  represents the thermal or kinetic pressure.
The fourth term on the  right-hand side has a purely quantum origin. The fifth term appears in the equation of motion (\ref{jjjjj}) through the
magnetization energy and represents the force of action of the external magnetic field on magnetic moments. The sixth
term on the  right-hand side describes a  force field produced by spin fluctuations.  The fourth and sixth terms are related and both of them are the generalization of the quantum Bohm potential appearing for the spin-1/2 particles.
This $spin$ $stress$ appears even in the absence of the electromagnetic fields and arises from the
inhomogeneity of the spin distribution \cite{12}. The second group of terms on the  right-hand side of (\ref{jjjjj}) gives a force field that represents
interactions between particles in the approximation of self-consistent field.

Equation (\ref{Mm})  is a generalization of the Bloch equation and  represents  the evolution of magnetic moment density. The first
term on the right-hand side of Eq. (\ref{Mm}) describes interaction of magnetic moments with the external magnetic field and with internal magnetic fields caused by the magnetic moments.  The second term on the  right-hand side of equation  (\ref{Mm})
describes the   $spin$  $torque$ influence on the magnetic moment density evolution which is written in the single
particle approximation.




\subsection{Energy evolution equation}

    Even in the self-consistent field approximation the set of equations
(\ref{nnn}) - (\ref{Mm}) is not closed, since it contains the thermal pressure $\wp$. We need to introduce the energy density and derived the spin and quantum terms in the energy density equation from the MPQH method.

The energy density taking into account the Coulomb and Spin-Spin interactions is given by \cite{1}, \cite{2} and \cite{112}
  \begin{widetext}
                                            $$
                                             \varepsilon(\textbf{r},t)=\int
                                             dR\sum^{N}_{j}\delta(\textbf{r}-\textbf{r}_{j})
                                             \frac{1}{4m_{j}}\{\psi^{+}_s\textbf{D}^{2}_{j}\psi_s+(\textbf{D}^{2}_{j}\psi_s)^{+}\psi_s\}(R,t)
                                              $$
                        \begin{equation} \label{Se}
                             \qquad\qquad\qquad   +\int dR\sum^{N}_{i\neq k}\delta(\textbf{r}-\textbf{r}_{j})
                               \frac{1}{2}\psi^{+}_s\{q_{j}q_{k}G_{jk}-\mu^{2}_{j}\sigma^{\alpha}_{j}
                               \sigma^{\beta}_{k}F^{\alpha\beta}_{jk}\}\psi_s(R,t).
                                \end{equation} \end{widetext}
              The first group of terms of Eq. (\ref{Se}) introduces the kinetic energy density and the second group of terms represents the potential energy density. The second group of terms consists of two parts. The first of them is the potential energy of the Coulomb interaction and the second one is the energy of Spin-Spin interaction. These terms are written via the Green functions described above.      Differentiation of (\ref{Se}) with respect to time and application of the Schrodinger equation with
Hamiltonian (\ref{H}) leads to the energy balance equation

\begin{widetext}  
 \begin{equation} \label{Se1}
           \frac{\partial}{\partial
           t}\varepsilon(\textbf{r},t)+\nabla\textbf{Q}
           (\textbf{r},t)=qj_{\alpha}(\textbf{r},t)E^{\alpha}_{ext}(\textbf{r},t)+
                                        \Im^{\alpha\beta}_M(\textbf{r},t)\partial_{\beta}B^{\alpha}_{ext}
                                        (\textbf{r},t)+A(\textbf{r},t),\end{equation} \end{widetext}                          where $A(\textbf{r},t)$   is the scalar field of work, $\Im^{\alpha\beta}_M$ represents the tensor of the magnetic moment flux density and $\textbf{Q}(\textbf{r},t)$  is the internal energy current. The first and second terms on the right-hand side of Eq. (\ref{Se1}) are caused by the interaction of charges with the external field and the magnetic moments with the external inhomogeneous magnetic field respectively. These terms are described the work of the external fields. The local work of the internal forces  $A(\textbf{r},t)$  has rather complicate form and in the self-consistent field approximation has a structure similar to the first two terms.

        The internal energy
                                      current density    is given by

   \begin{widetext}
$$                                      \textbf{Q}(\textbf{r},t)=\int
dR\sum^{N}_{j}\delta(\textbf{r}-\textbf{r}_{j})
\frac{1}{8m^2_{j}}\Biggl(\psi^{+}_s\textbf{D}_{j}\textbf{D}^{2}_{j}\psi_s
+(\textbf{D}_{j}\textbf{D}^{2}_{j}\psi_s)^{+}\psi_s+\textbf{D}^{+}_{j}\psi^{+}_s\textbf{D}^{2}_{j}\psi_s +
(\textbf{D}^{2}_{j}\psi_s)^{+}\textbf{D}_{j}\psi_s\Biggr)
$$

      \begin{equation}\label{SQ}
\qquad\qquad\qquad\qquad+\int dR\sum^{N}_{j\neq k}\delta(\textbf{r}-\textbf{r}_{j})\frac{1}{4m_{j}}
\Biggl(\psi^{+}_s(R,t)(q_{j}q_{k}G_{jk}-\mu^{2}_{j}\sigma^{\alpha}_{j}\sigma^{\beta}_{k}F^{\alpha\beta}_{jk})\textbf{D}_{j}\psi_s(R,t)+k.c.\Biggr) \end{equation}    \end{widetext}
The full current of energy or energy flux (\ref{SQ}) consists of  the kinetic  energy current density, which presented by the first group of terms and potential energy current density, which is described by the second group of terms.

        The  scalar field of work  $A(\textbf{r},t)=A_{s-s}(\textbf{r},t)+A_{cl}(\textbf{r},t)+A_{s}(\textbf{r},t)$  in (\ref{Se1}) consists of the Coulomb force density $A_{cl}(\textbf{r},t)$ and Spin-Spin force density $A_{s-s}(\textbf{r},t)$.

         After the presentation of the wave function in the exponential form (\ref{psi}), using the Madelung decomposition (\ref{z4}), the tensor
of the magnetic moment flux density takes the form \cite{12}

                       \begin{equation}  \label{Jj}
\Im^{\alpha\beta}_{M}(\textbf{r},t)=M^{\alpha}\upsilon^{\beta}(\textbf{r},t)+\gamma_s^{\alpha\beta}(\textbf{r},t)+d_T^{\alpha\beta}(\textbf{r},t), \end{equation} where    the spin contribution to the magnetization current takes the form

      $$  \gamma^{\alpha\beta}_s(\textbf{r},t)=-\sum_{S}\int
dR\sum^{N}_{j=1}\delta(\textbf{r}-\textbf{r}_{j})\frac{2\mu_j\varepsilon^{\alpha\mu\nu}}{m_j\hbar}\times
        $$ \begin{equation}  \label{JM5}  \qquad\qquad\qquad\qquad\qquad\qquad\qquad \times a^{2}(R,t)s_{j}^{\mu}\nabla_{j}^{\beta}s_{j}^{\nu}
                                                            \end{equation}
                                                          where $s_j$ represents the particle spin
                                                             and the thermal  spin fluctuations takes the form  \cite{24}

   \[ \label{ksi} \xi^{\alpha}_{j}(\textbf{r},R,t)=s^{\alpha}_{j}(R,t)-s^{\alpha}(\textbf{r},t), \] In the context of quantum hydrodynamics  the second term in Eq. (\ref{Jj}) describes  the additional \emph{spin torque} and its explicit form to be

                            \begin{equation}  \label{JM5}
\gamma^{\alpha\beta}_s(\textbf{r},t)=-\frac{\hbar}{2m\mu_j}
\varepsilon^{\alpha\gamma\lambda}M_{\gamma}(\textbf{r},t)\partial_{\beta}
(\frac{M^{\lambda}(\textbf{r},t)}{n(\textbf{r},t)})+\Theta^{\alpha\beta}_s(\textbf{r},t).
\end{equation}
The  quantum equivalent of the thermal speed contributes to the magnetization current $d^{\alpha\beta}_T$

\begin{equation} \label{sd}  d_T^{\alpha\beta}(\textbf{r},t)=\sum_{S}\int
  dR\sum^{N}_{j=1}\delta(\textbf{r}-\textbf{r}_{j})\frac{2\mu_j}{\hbar}a^{2}(R,t)u^{\beta}_js^{\alpha}_j.
                                                                                       \end{equation}
      The spin density thermal fluctuations  are presented  in the form of \cite{24}

          $$\Theta_s^{\alpha\beta}(\textbf{r},t)=-
 \varepsilon^{\alpha\mu\nu}\sum_{S}\int
dR\sum^{N}_{j=1}\delta(\textbf{r}-\textbf{r}_{j})\frac{2\mu_j}{m_j\hbar}\times
                                           $$\begin{equation} \label{sd1}\qquad\qquad\qquad\qquad\qquad\qquad\times a^{2}(R,t)\xi_{j\mu}\nabla_j^{\beta}\xi^{\nu}_j.\end{equation}
                           Using
the fact that the velocity field  is determined by  (\ref{z4}) and using the definition  (\ref{Jj}) we derive the thermal energy $\epsilon$ evolution equation

    \begin{widetext}
     $$
 n(\frac{\partial}{\partial
 t}+\textbf{v}\nabla)\epsilon+\nabla\textbf{q}+\frac{\hbar^{2}}{4mn}\{\partial_{\alpha}n\}\{\partial^{\beta}n\}\partial_{\beta}\upsilon^{\alpha}
                                                                                                                                                                       $$

 $$ -\frac{\hbar^{2}}{4m}\partial_{\alpha}(n\partial_{\alpha}\partial^{\beta}\upsilon^{\beta})
 -\frac{\hbar^{2}}{4m}\partial_{\alpha}\partial^{\beta}n\partial_{\beta}\upsilon^{\alpha}-\frac{\hbar^2M_{\gamma}}{4m\mu^2}\partial_{\alpha}\partial^{\beta}(\frac{M^{\gamma}(\textbf{r},t)}{n})\partial_{\beta}\upsilon^{\alpha}
 $$

$$\qquad\qquad\qquad
+Q^{\alpha\beta}_s\nabla_{\beta}\upsilon^{\alpha}+p^{\alpha\beta}\partial_{\beta}\upsilon_{\alpha}=d_T^{\alpha\beta}\partial_{\beta}B^{\alpha}_{ext}
+\nabla_{\beta}B^{\alpha}_{ext}\Theta^{\alpha\beta}_s
$$

\begin{equation}   \label{Se3}\qquad\qquad\qquad\qquad\qquad\qquad\qquad\qquad
-\frac{\hbar}{2m\mu}\varepsilon^{\alpha\mu\nu}M_{\mu}\partial_{\beta}(\frac{M^{\nu}}{n})\nabla_{\beta}B^{\alpha}_{ext}.\end{equation}
               \end{widetext}Let us discuss the physical significance of the terms on the left-hand side of the energy equation  (\ref{Se3}). The third, fourth  and fifth terms on the left-hand side in Eq.  (\ref{Se3}) describe a quantum force produced by Bohm potential. The eighth term represents   the well
known pressure tensor influence. The sixth term on the left-hand side  of (\ref{Se3}) characterizes the $spin$ $stress$ in influence on the energy density evolution and the third term on the right-hand side describes the $spin$ $torque$ influence.   The seventh term  on the left-hand side, the first and second terms on the right-hand side describe the thermal effects. The forth term  on the right-hand side in Eq.  (\ref{Se3}) is the total force field density.

     The specific  "heat" or thermal energy density   contains the
thermal, quantum and spin contributions and takes the form

         \begin{widetext}
            $$   
  n\epsilon(\textbf{r},t)=\int
  dR\sum^{N}_{j}\delta(\textbf{r}-\textbf{r}_{j})a^2(R,t)(\frac{m_j\textbf{u}^2_j}{2}-
  \frac{\hbar^2}{2m_j}\frac{\triangle_ja}{a}+\frac{1}{2m_j}|\nabla_{\alpha}s^{\alpha}_j|^2)
  $$
  \begin{equation} \label{Se2}  \qquad\qquad+ \frac{q^2}{2}\int
  d\textbf{r}^{'}G(\textbf{r},\textbf{r}^{'})n_2(\textbf{r},\textbf{r}^{'},t)- \frac{1}{2}\int
  d\textbf{r}^{'}F^{\alpha\beta}(\textbf{r},\textbf{r}^{'})M^{\alpha\beta}(\textbf{r},\textbf{r}^{'},t).
  \end{equation}\end{widetext}  We define the thermal energy density via the amplitude of the many-particle wave function $a(R,t)$ and thermal velocity of particles $\textbf{u}_j$.    The first term on the right-hand side of the expression  (\ref{Se2})  describes the quantum equivalent of the thermal speed contribution, the second term   characterizes the quantum Madelung potential   contribution and the third term presents the \emph{internal spin potential} influence. The forth and fifth terms are the potential energy density of the particle interaction, namely the Coulomb interaction of charges and
spin-spin interactions. The \emph{internal spin potential} can be rewritten in the approximation of noninteracting particles using the definition (\ref{ksi})

                 \begin{widetext}
  $$ n\epsilon(\textbf{r},t)=\int
  dR\sum^{N}_{j}\delta(\textbf{r}-\textbf{r}_{j})\frac{a^2(R,t)}{2m_j}|\nabla_{\alpha}s^{\alpha}_j|^2=\frac{n(\textbf{r},t)}{2m}|\nabla_{\alpha}s^{\alpha}(\textbf{r},t)|^2
 $$

 \begin{equation} \label{nee}\qquad\qquad\qquad\qquad\qquad\qquad+\int
  dR\sum^{N}_{j}\delta(\textbf{r}-\textbf{r}_{j})a^2(R,t)\frac{1}{2m_j}|\nabla_{\alpha}\xi^{\alpha}_j|^2.\end{equation}\end{widetext}
The  energy current  takes the form
                  \begin{widetext}
      $$ q^{\alpha}(\textbf{r},t)=\int
                      dR\sum^{N}_{j}\delta(\textbf{r}-\textbf{r}_{j})a^2(R,t)\Biggl(u^{\alpha}_j(\frac{m_j\textbf{u}^2_j}{2}-\frac{\hbar^2}{2m_j}\frac{\triangle_ja}{a}+\frac{1}{2m_j}|\nabla_{\alpha}s^{\alpha}_j|^2)
                      $$ \begin{equation} \label{q2}\qquad\qquad\qquad\qquad-\frac{\hbar^2}{2m_{j}}\frac{\partial}{\partial
                      x^{\alpha}_{j}}(u_{j\beta}\frac{\partial\ln a}{\partial x^{\beta}_{j}})-
                      \frac{\hbar^2}{4m_{j}} \frac{\partial^2u_{j\beta}}{\partial
                      x^{\alpha}_{j}\partial
                      x^{\beta}_{j}}+\frac{u_{j\beta}}{m_j}\partial^{\alpha}s_{j\gamma}\partial^{\beta}s_j^{\gamma}
                      \Biggr).
               \end{equation}     \end{widetext}
The first term in Eq. (\ref{q2})  represents the thermal energy current of kinetic energy which is proportional to $\textbf{u}^2_j$, the second, fourth and fifth terms represent the
quantum energy current which is proportional to $\hbar^2$. The fourth and fifth terms are quantum-thermal terms since they are contain the Plank constant
and the thermal velocities  $\textbf{u}_j$.
 The third and sixth terms represent  the
spin  contribution to the energy current. We can see that the definition (\ref{q2})  contains information
about the spin effects.  Spin contributions can be represent in the form of
                    
          \begin{widetext}
             $$
             \int dR\sum^{N}_{j}\delta(\textbf{r}-\textbf{r}_{j})a^2\frac{u^{\alpha}_j}{2m_j}|\nabla_{\alpha}s^{\alpha}_j|^2
             $$

             \begin{equation}\label{qq} \qquad\qquad\qquad\qquad\qquad\qquad=
             \int dR\sum^{N}_{j}\delta(\textbf{r}-\textbf{r}_{j})a^2\frac{u^{\alpha}_j}{2m_j}\Biggl(2\nabla s\cdot\nabla\xi_j+|\nabla\xi_j|^2 \Biggr)
              \end{equation}

       and

            $$ \int dR\sum^{N}_{j}\delta(\textbf{r}-\textbf{r}_{j})a^2\frac{u^{\alpha}_j}{m_j}\partial^{\alpha}s_{j\gamma}\partial^{\beta}s_j^{\gamma}
              =
             \int dR\sum^{N}_{j}\delta(\textbf{r}-\textbf{r}_{j})a^2\frac{u^{\alpha}_j}{m_j}\partial^{\alpha}\xi_{j\gamma}\partial^{\beta}\xi_j^{\gamma}
               $$

               \begin{equation}\label{q}\qquad\qquad\qquad\qquad\qquad +\int dR\sum^{N}_{j}\delta(\textbf{r}-\textbf{r}_{j})a^2\frac{u^{\alpha}_j}{m_j}\Biggl(
           \partial^{\alpha}s_{j\gamma}\partial^{\beta}\xi_j^{\gamma}+\partial^{\alpha}\xi_{j\gamma}\partial^{\beta}s_j^{\gamma}   \Biggr)  \end{equation} \end{widetext}
              The definitions represent the spin energy density and spin energy current respectively.

     \section{Conclusions}

      In this paper we analyzed  energy  evolution caused by the
magnetic moment density dynamics in systems of charged  $ 1/2-spin$ particles.    We consider the $Coulomb$ and $Spin-Spin$
interactions in the equations. The system of MQHD equations we have constructed comprises equations of continuity (\ref{nnn}),
of the momentum balance (\ref{jjjjj}),  equation
for the energy evolution (\ref{Se3}), equation
for the magnetization evolution (\ref{Mm}). We have derived the  spin part of quantum Bohm potential in
the equations we are interested in, determining the system dynamics, are the hydrodynamic equations for the spinning plasma. This equations (\ref{jjjjj} and \ref{Mm})
have an additional  new spin-dependent  $spin$ $stress$  and  $spin$  $torque$ which have been derived taken into account thermal fluctuations of  spins about the macroscopic average.

We have derived the energy evolution equation. The main objective of this paper was
to construct  a  generalized  thermal energy density   equation (\ref{Se3}) for  spinning quantum plasmas that contains spin contributions.  We define the thermal energy density via the amplitude of the many-particle wave function $a(R,t)$ and thermal velocity of particles $\textbf{u}_j$ (\ref{Se2}). We have found the spin part of thermal energy density or the \emph{internal spin potential} which can be rewritten in the approximation of noninteracting particles using the definition  (\ref{nee}). We have derived the spin  contribution to the kinetic energy current  (\ref{qq}) and (\ref{q}).
 It was showed that in the absence of external fields the dynamics of energy density  is subject to the quantum Bohm potential and and spin part of quantum Bohm potential or $spin$ $stress$. We have described the spin contributions in the energy evolution for the point-like charged particles, while considering ions we deal with the finite size objects. It is necessary  to include finite size of particles. Corresponding model was developed  in Ref. \cite{3000}.

        \end{document}